\documentclass[a4paper,aps,prd,showpacs,preprintnumbers,twocolumn,nofootinbib]{revtex4}

\usepackage[normalem]{ulem} 
\usepackage{graphicx}
\usepackage{amsmath}
\usepackage{amsfonts}   
\usepackage{amssymb}  
\usepackage{mathrsfs}
\usepackage{epsfig,bbm}
\usepackage{bm}
\usepackage{color}
\usepackage{dsfont}

\newcommand{\troisj}[6]{\left(\begin{array}{ccc}
      #1 & #2 & #3 \\
      #4 & #5 & #6\end{array}\right)}
\begin{document}
\newcommand{\average}[1]{\langle{#1}\rangle_{{\cal D}}}
\newcommand{\dd}{{\rm d}}
\newcommand{\snumb}{{\cal N}}
\newcommand{\etal}{{\it et al.}}
\newcommand{\obs}{{\rm o}}

\title{Weak-lensing $B$-modes as a probe of the isotropy of the universe}

\author{Thiago S. Pereira}
\email{tspereira@uel.br}
\affiliation{Departamento de F\'isica, Universidade Estadual de Londrina,
 86051-990, Londrina, Paran\'a, Brazil.}

\author{Cyril Pitrou}
\email{pitrou@iap.fr}
 \affiliation{
             Institut d'Astrophysique de Paris,
             Universit\'e Pierre~\&~Marie Curie - Paris VI,
             CNRS-UMR 7095, 98 bis, Bd Arago, 75014 Paris, France;\\
             Sorbonne Universit\'es, Institut Lagrange de Paris,
             98 bis, boulevard Arago, 75014 Paris, France}
             
\author{Jean-Philippe Uzan}
\email{uzan@iap.fr}
 \affiliation{
             Institut d'Astrophysique de Paris,
             Universit\'e Pierre~\&~Marie Curie - Paris VI,
             CNRS-UMR 7095, 98 bis, Bd Arago, 75014 Paris, France;\\
             Sorbonne Universit\'es, Institut Lagrange de Paris,
             98 bis, boulevard Arago, 75014 Paris, France}

\begin{abstract}
We compute the angular power spectrum of the $B$-modes of the  weak-lensing shear in a spatially anisotropic spacetime. We find that there must also exist off-diagonal correlations between the $E$-modes, $B$-modes, and convergence that allow one to reconstruct the eigendirections of expansion. Focusing on future surveys such as Euclid and SKA, we advertise that observations can constrain the geometrical shear in units of the Hubble rate at the percent level, or even better, offering a new and powerful method to probe our cosmological model. However, the power of this new technique still requires further investigations and a full analysis of signal-to-noise ratio. We hope that it can provide a new tool for precision cosmology.
\end{abstract}

\maketitle

According to the standard lore~\cite{slore}, in a homogeneous and isotropic background spacetime, weak-lensing by the large scale structure of the universe induces a shear field which, to leading order, only contains $E$-modes. The level of $B$-modes is used as an important sanity check during the data processing. On small scales, $B$-modes arise from intrinsic alignments~\cite{intrinsic}, Born correction, lens-lens coupling~\cite{lenslens}, and gravitational lensing due to the redshift clustering of source galaxies~\cite{galcluster}. On large angular scales in which the linear regime holds, it was demonstrated~\cite{pup} that non-vanishing $B$-modes would be a signature of a deviation from the isotropy of the expansion, these modes being generated by the coupling of the background Weyl tensor to the $E$-modes.

In this letter, we emphasize that, as soon as local isotropy does not hold at the background level, there exist a series of weak-lensing observables that allow one to fully reconstruct the background shear and thus test isotropy. We also quantify their magnitude for typical surveys such as Euclid~\cite{euclid} and SKA~\cite{ska}. As a consequence of the existence of $B$-modes, it can be demonstrated that: (1) the angular correlation function of the $B$-modes, $C_\ell^{BB}$, is non-vanishing~\cite{pup}; (2) they also correlate with both the $E$-modes and the convergence, leading to the off-diagonal cross-correlations $\langle B_{\ell m}E^{\star}_{\ell\pm1 \,m-M}\rangle$ and  $\langle B_{\ell m}\kappa^{\star}_{\ell\pm1 \,m-M}\rangle$ in which $E_{\ell m}$ and $B_{\ell m}$ are the components of the decomposition of the $E$- and $B$-modes of the cosmic shear in (spin-2) spherical harmonics, and $\kappa_{\ell m}$ are the components of the decomposition of the convergence in spherical harmonics~\cite{newppu}; (3) the deviation from isotropy also generates off-diagonal correlations between $\kappa$ and $E$-modes, $\langle E_{\ell m} {E}^{\star}_{\ell\pm2 \,m-M}\rangle$,  $\langle{\kappa}^X_{\ell m} {\kappa}^{X\,\star}_{\ell\pm2 \,m-M}\rangle$, and $\langle{E}^X_{\ell m} {\kappa}^{X\,\star}_{\ell\pm2 \,m-M}\rangle$.

Our companion article~\cite{newppu} provides all the technical details of the theoretical computation of these correlators. In this letter, we estimate the information that can be extracted from weak-lensing by focusing on these correlations, and illustrate its power to constrain a late time anisotropy.\\

We assume that the background spacetime is spatially flat and homogeneous, but enjoys an anisotropic expansion. It can be described by a Bianchi~I universe with metric
\begin{eqnarray}
\dd s^2&=&-\dd t^2+a(t)^2\gamma_{ij}(t)\dd x^i\dd x^j\,, \label{e:metric2}
\end{eqnarray}
where $a(t)$ is the volume averaged scale factor. The spatial metric $\gamma_{ij}$ is decomposed as $\gamma_{ij}(t)=\exp[2\beta_{i}(t)]\delta_{ij}$ with the constraint $\sum_{i=1}^3\beta_i=0$. The geometrical shear, not to be confused with the cosmic shear, is defined as
\begin{equation}\label{e:decbeta}
\sigma_{ij}\equiv\frac{1}{2}\dot\gamma_{ij}\,.
\end{equation}
Its amplitude, $\sigma^2 \equiv \sigma_{ij}\sigma^{ij} = \sum_{i=1}^3\dot\beta_i^2$, characterizes the deviation from a Friedmann-Lema\^{\i}tre spacetime. We also define $H=\dot a/a$.

At this stage, it is important to stress that, since $\sigma_{ij}$ is traceless, it has 5 independent components. In the limit in which $\sigma/H\ll~1$ (the relevant limit to constrain small departures from isotropic expansion) each of the 5 correlators -- $\langle B_{\ell m}E^{\star}_{\ell\pm1 \,m-M}\rangle$, $\langle B_{\ell m}\kappa^{\star}_{\ell\pm1 \,m-M}\rangle$, $\langle E_{\ell m} {E}^{\star}_{\ell\pm2 \,m-M}\rangle$,  $\langle{\kappa}^X_{\ell m} {\kappa}^{X\,\star}_{\ell\pm2 \,m-M}\rangle$, and $\langle{E}^X_{\ell m} {\kappa}^{X\,\star}_{\ell\pm2 \,m-M}\rangle$ -- is of first order in $\sigma/H$ and has five independent components ($M=-2\ldots+2$) that allow one, in principle, to reconstruct the independent components of $\sigma_{ij}$. The angular power spectrum $C_\ell^{BB}$, on the other hand, scales as $(\sigma/H)^2$ and, while it can point to a deviation from isotropy, it does not allow one to reconstruct the principal axis of expansion.

Following our earlier works~\cite{pup,newppu}, we adopt an observer based point of view, i.e. we compute all observable quantities in terms of the direction of observation ${\bm n}_\obs$. The main steps of the computation are the resolution of the background geodesic esquation (which provides the local direction ${\bm n}({\bm n}_\obs,t)$ on the lightcone and hence the definition of the local Sachs basis), the resolution of the Sachs equation at the background level and at linear order in perturbation, and a multipole decomposition of all the quantities, a step more difficult than usual because of the fact that ${\bm n}\not={\bm n}_\obs$. We then perform a small shear limit in which one can isolate the dominant, followed by the use of the Limber approximation (although not mandatory). This provides the expressions of the different correlators
\begin{eqnarray}
{}^{XZ}{\cal A}^{M}_{\ell_1\,\ell_2}  &\equiv&  \sum_{m}\sqrt{5}(-1)^{m+\ell_1+\ell_2}\troisj{\ell_1}{2}{\ell_2}{-m}{M}{m-M}\nonumber\\
&&\qquad\times \langle{X}^{X}_{\ell_1 m}Z^{X\,\star}_{\ell_2,m-M}\rangle\,
\end{eqnarray}
that take the form (see Eqs. (7.17-7.19) of Ref.~\citep{newppu})
\begin{eqnarray}
{}^{BE}{\cal A}^{M}_{\ell\ell\pm1}  &=&  {\rm i} \frac{{}_2F_{\ell2\ell\pm1}}{\sqrt{5}}\,\,  {\cal P}_{\ell\pm1 M}^{EE} \,,\nonumber\\
{}^{B\kappa}{\cal A}^{2M}_{\ell\ell\pm1}  &=&  {\rm i} \frac{F_{\ell2\ell\pm1}}{\sqrt{5}}\,\,  {\cal P}_{\ell\pm1 M}^{E\kappa } \,,\nonumber\\
{}^{EE}{\cal A}^{M}_{\ell\ell\pm2}  &=& \frac{{}_2F_{\ell2\ell\pm2}}{\sqrt{5}}\,\,  \,\left({\cal P}_{\ell\pm2 M}^{EE}+ {\cal P}_{\ell\,M}^{EE}\right)\,,\nonumber\\
{}^{\kappa \kappa}{\cal A}^{M}_{\ell\ell\pm2}  &=& \frac{F_{\ell2\ell\pm2}}{\sqrt{5}}\,\,  \,\left({\cal P}_{\ell\pm2 M}^{\kappa \kappa}+ {\cal P}_{\ell\,M}^{\kappa \kappa}\right)\,,\nonumber\\
{}^{E\kappa}{\cal A}^{M}_{\ell\ell\pm2}  &=& \frac{{}_2F_{\ell2\ell\pm2}}{\sqrt{5}}\,\,  {\cal P}_{\ell\pm2 M}^{E\kappa}+\frac{F_{\ell2\ell\pm2}}{\sqrt{5}}\,\, {\cal P}_{\ell\,M}^{\kappa E}\,.\nonumber
\end{eqnarray}
${}_sF_{\ell_12\ell_2}$ are explicit functions of the multipoles given in Appendix~D.4 of Ref.~\cite{newppu} and defined in Ref.~\cite{Flhu}. The general form of the quantities ${\cal P}^{XY}_{\ell m}$ is given by Eq. (7.14) of Ref.~\cite{newppu} and, in the Limber approximation, they reduce to
\begin{equation}
\left[\begin{array}{c}
{\cal P}_{\ell\,M}^{\kappa \kappa}\\ {\cal P}_{\ell\,M}^{\kappa E}\\ {\cal P}_{\ell\,M}^{EE}
\end{array}\right]
=\frac{1}{4}\left[\begin{array}{c} \ell^2(\ell+1)^2\\ \ell(\ell+1)\sqrt{\frac{(\ell+2)!}{(\ell-2)!}}\\ \frac{(\ell+2)!}{(\ell-2)!}\end{array}\right]
\times {\cal P}_{\ell\,M}\,,
\end{equation}
with (see Eqs. (7.20) and (7.21) of Ref.~\cite{newppu})
\begin{eqnarray}
{\cal P}_{\ell\,M} &\equiv& \int_0^\infty
\frac{\dd \tilde \chi}{\tilde \chi^2}P\left(\frac{L}{\tilde \chi}\right)\alpha_{2 M}(\tilde \chi)\nonumber\\
&&\times \left|T^\varphi\left(\frac{L}{\tilde \chi},\tilde \chi\right)\int_{\tilde \chi}^\infty \dd\chi \right. \left.\snumb(\chi)\frac{(\chi-\tilde \chi)}{\chi\tilde \chi} \right|^2\,,
\end{eqnarray}
and $L\equiv \ell+1/2$. $P(k)$ stands for the primordial power spectrum of the metric fluctuations, $T^\varphi\left(x,\eta\right)$ is the transfer function of the deflecting potential given, as usual, by the sum of the two Bardeen potentials. They are both evaluated on the past lightcone parametrized by the radial coordinate $\chi$; in the Limber approximation $k=L/\tilde\chi$. $\snumb(\chi)$ is the source distribution and depends on the survey. $\alpha_{\ell m}(\chi)$ is the multipolar coefficient of the deflection angle $\alpha({\bm n}_\obs,\chi)$ expanded in spherical harmonics. At the lowest order in $\sigma/H$, only its $\ell=2$ components are non-vanishing (see \S VII.B.2 of \cite{newppu} for the expressions for $\alpha_{2m}$).

\begin{figure}[ht!]
\centering
\includegraphics[width=7cm]{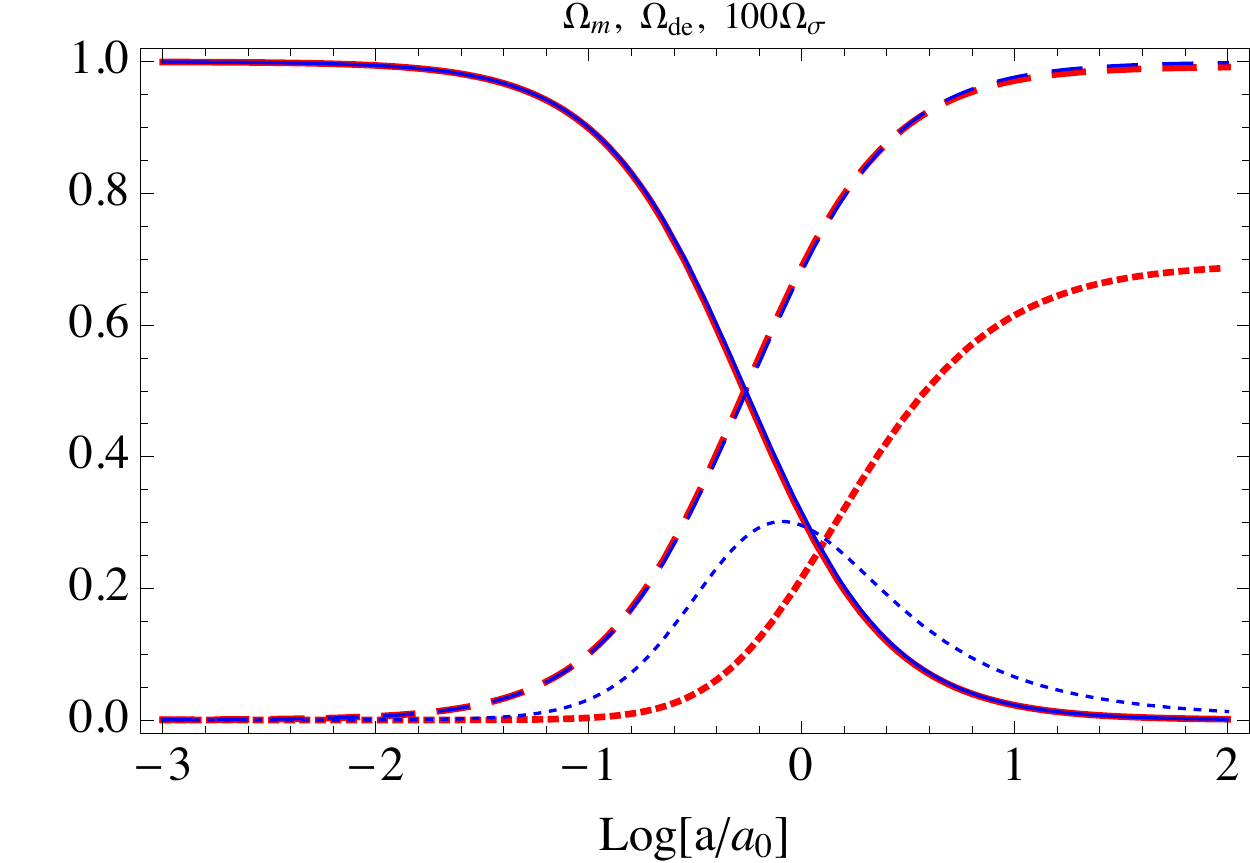}
\centering
\caption{Time evolution of models A (red)
 and B (blue).
The plot shows the contribution to the expansion of matter (solid line), dark energy (dashed line) and geometrical shear (dotted line, and magnified by a factor 100).}
\label{Fig1}
\end{figure}

While the previous off-diagonal correlators are the most direct consequence of a late time anisotropy, most experiments are designed to measure the angular power spectrum. 
\begin{widetext}
We obtain~\cite{newppu} that for the $B$-modes
\begin{eqnarray}
C_\ell^{BB} = \frac{2}{5\pi}\int_0^\infty k^2 \dd k P(k) \sum_{s=    \pm 1}\frac{({}_2
   F_{\ell\,2\,\ell+s})^2}{2 \ell+1}  \sum_m \left|\int_0^\infty \dd
   \chi \snumb(\chi) \int_0^\chi \dd \chi' \alpha_{2 m}(\chi')  g^E_{\ell+s}(k,\chi,\chi') \right|^2\,,
\end{eqnarray}
where 
${}_2 F_{\ell\,2\,\ell+s}$ is a function of $\ell$ and the functions $g^E_\ell$ are expressed in terms of spherical Bessel functions and given by Eq. (6.44) of 
Ref.~\cite{newppu}.\\
\end{widetext}

To estimate these correlators, we need to go through the following steps. First, we need to solve the geodesic equation for the background spacetime in order to determine ${\bm n}({\bm n}_0,\chi)$ and the deflection angle. We then need to describe and solve the evolution of metric perturbations (in order to determine the transfer function $T^\varphi$ of the lensing potential.) \\

During inflation, the spacetime isotropizes, letting only tiny, if any, signatures on the cosmic microwave background (CMB)~\cite{early-aniso}, which has been constrained observationally~\cite{cmb-aniso,cmb-aniso2}. On the other hand, many models of the dark sector~\cite{demodels} have considered the possibility that dark energy enjoys an anisotropic stress. This is a generic prediction of bigravity~\cite{bigrav} and backreaction~\cite{backreac}. This has stimulated the investigation of methods to constrain a late time anisotropy, using e.g. the integrated Sachs-Wolfe effect~\cite{cmb-aniso2}, large scale structure and the Hubble diagram of supernovae in different fields~\cite{Fleury2014,observation}.

From a phenomenological point of view, one can consider a dark energy sector with an anisotropic stress. Its stress-energy tensor is then decomposed as $T^\mu_\nu=(\rho+P)u^\mu u_\nu+P \delta^\mu_\nu+\Pi^\mu_\nu$ where the anisotropic stress tensor $\Pi^\mu_\nu$ is traceless ($\Pi^\mu_\mu=0$), transverse ($u_\mu\Pi^\mu_\nu=0$), and has 5 degrees of freedom encoded in its spatial part $\Pi^i_j$. It can be decomposed in terms of an anisotropic equation of state~\cite{w-anisotrope,limit-dw} as $P_i^j = \rho_{\rm de}\left(w\delta_i^j + \Delta w_i^j\right)$. Here, $w$ is the usual equation of state (we assume $w=-1$ as for a cosmological constant) and we need to model $\Pi^i_j$. 

The background equations then  take the form
\begin{eqnarray}
3H^2&=&\kappa(\rho_{\rm m}+\rho_{\rm de})+\frac{1}{2}\sigma^2\,,\label{g00} \\
(\sigma^i_j)^\cdot &=& -3H\sigma^i_j + \kappa\Pi^i_j\,. \label{gTT}  \\
\dot\rho_{\rm m} &=& -3H\rho_{\rm m}\,, \label{e:dT_mat} \\
\dot\rho_{\rm de}&=& -\sigma_{ij}\Pi^{ij} \label{e:dT_de} \,.
\end{eqnarray}
The first equation is the Friedmann equation, the second is obtained from the traceless and transverse part of the Einstein equation and dictates the evolution of the shear. The last two equations are the continuity equations for the dark matter ($P=\Pi^i_j=0$) and dark energy sectors.

Simple models can be  built by phenomenologically relating $\Pi^i_j$ to the geometrical shear as $\Pi^i_j = \lambda \sigma^i_j \equiv  \sigma^i_j/\kappa\tau_{_\Pi}$, where $\tau_{_\Pi}$ can be time-dependent. When it is constant, the shear grows exponentially as $\sigma^i_j = B^i_j\left(\frac{a_0}{a}\right)^3\hbox{e}^{t/\tau_{_\Pi}}$. Since $\sigma_{ij}$ is small today, there is some fine tuning. We thus consider two classes of models defined by
\begin{equation}\label{e:dec_Pi}
(A):\quad \Pi^i_j\equiv\rho_{\rm de}\Delta w^i_j\,;\qquad
(B):\quad \Pi^i_j\equiv g(a)\Delta w^i_j.
\end{equation} 
This assumes that the anisotropic stress evolves with time while keeping its eigenvalues in a constant ratio. The function $g(a)$ is arbitrary, and we note that when $g(a)=3H/H_0$, $\sigma^i_j=\mathcal{C}^i_j\left(\frac{a_0}{a}\right)^3+\kappa\frac{{\Delta w}_{j}^{i}}{H_0}$, so that at late time $\sigma^2\propto \kappa^2{\Delta w}^2/H_0^2$ is constant. In the models (A), the dark energy triggers the anisotropic phase. It was argued~\cite{limit-dw} that next generation surveys will be capable of constraining anisotropies at the 5\% level in terms of the anisotropic equation of state, a number to keep in mind for comparison with weak-lensing. Eq.~(\ref{gTT}) implies that 
$$
\sigma^i_j=\left(\frac{a_0}{a}\right)^3\left[\mathcal{C}^i_j
+\kappa\int\Pi^i_j\left(\frac{a}{a_0}\right)^2\frac{d(a/a_0)}{H}\right],
$$
while Eq.~(\ref{e:dT_de}) implies that $\rho_{\rm de}$ will decrease as
$$
 \rho_{\rm de}=\rho_{\rm de0} \exp\left[- \int\sigma^j_i\Delta w^i_j\frac{da}{aH} \right].
$$
Figure~\ref{Fig1} depicts the evolution of the density parameters for a model of each class.

To evaluate the angular power spectrum of the $E$- and $B$-modes, one needs to specify $\snumb(\chi)$. For this we consider the distributions of the future Euclid and SKA experiments. The normalised Euclid redshift distribution~\cite{Beynon:2009yd,euclid} is
\begin{equation}
\snumb(z)=Az^{2}\exp\left[-\left(\frac{z}{z_{0}}\right)^{\beta}\right]
\end{equation}
with $A=5.792$, $\beta=1.5$ and $z_{0}=0.64$. For SKA, we use the SKA Simulated Skies simulations~\cite{Wilman:2008ew} of radio source population using the extragalactic radio continuum sources in the central $10\times10$ sq. degrees up to $z=20$. The SKA normalized redshift distribution is~\cite{Andrianomena:2014sya} 
\begin{equation}
\snumb(z)=A\frac{z^n}{(1+z)^m}\exp\left[-\frac{(a+bz)^{2}}{(1+z)^2}\right]\,, \qquad z<20
\end{equation}   
with best fit parameters  $a = -1.806, b = 0.388, m = 2.482, n = 0.838$ and $A=1.610$, which give a description accurate to the percent level.

\begin{figure}[ht!]
\centering
\includegraphics[width=7cm]{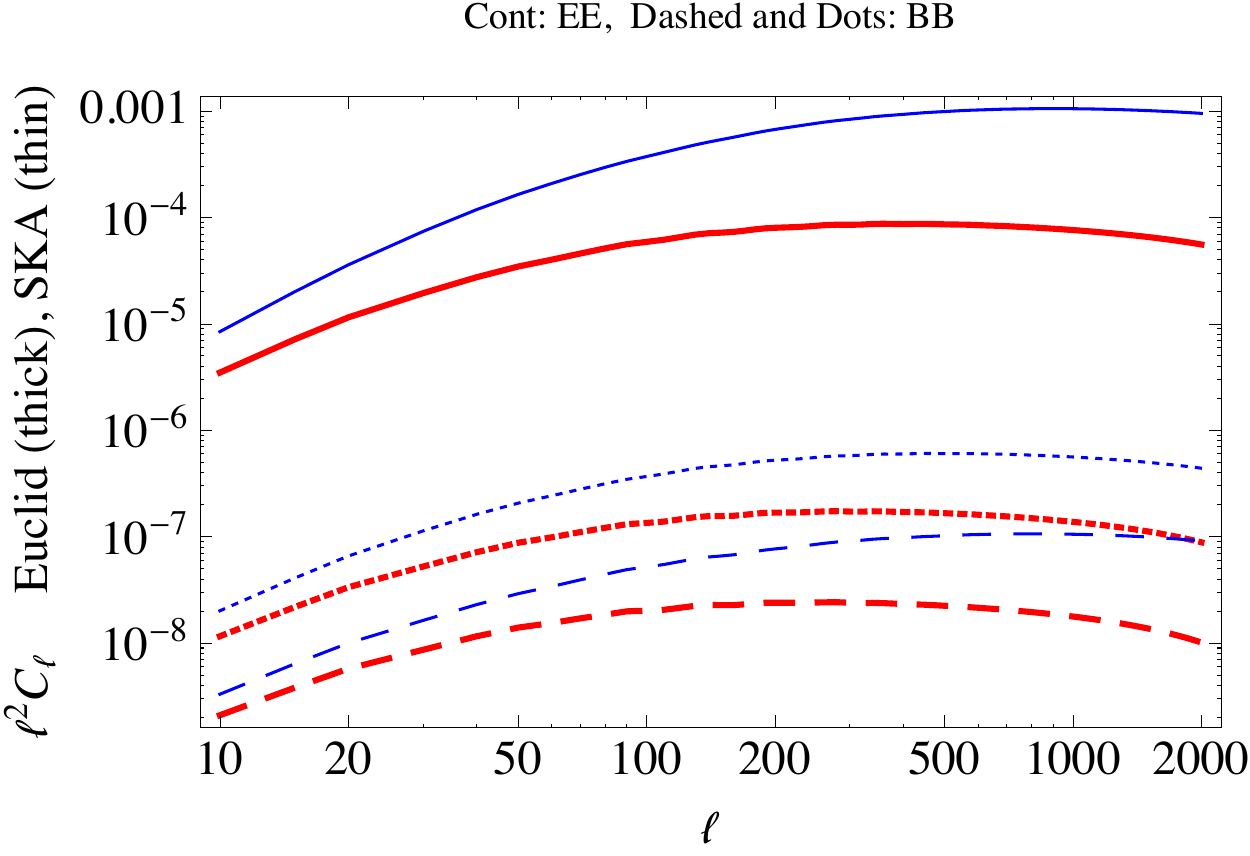}
\centering
\caption{Angular power spectra of the $E$- and $B$-modes (resp. solid and dashed lines) for the Euclid (red lines) 
and SKA (blue lines)
 surveys for models A (long dashed) and B (short dashed).}
\label{Fig2}
\end{figure}

\begin{figure}[ht!]
\centering
\includegraphics[width=7cm]{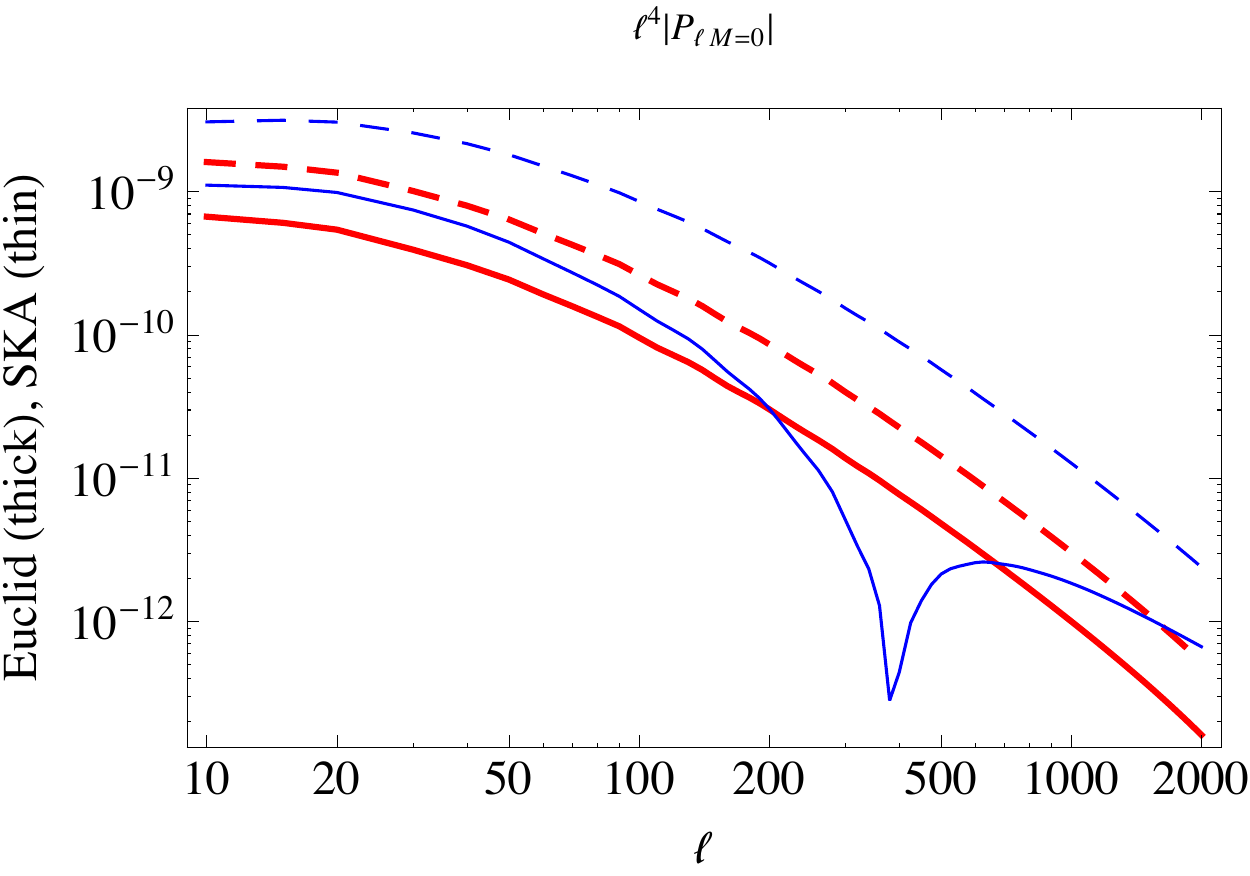}
\centering
\caption{The five correlators ${}^{XY}{\cal A}^{M}_{\ell\ell\pm1}$ are, up to shape factors, proportional to $\ell^4{\cal P}_{\ell M}$. We plot the latter for the models A (dashed) and B (solid) for the Euclid (red) and SKA (blue). For clarity we plot only the $M=0$ component.}
\label{Fig3}
\end{figure}

Figure~\ref{Fig2} depicts the two angular power spectra for these two surveys. In the linear regime the $B$-mode contribution is expected to vanish and the terms $\ell^4{\cal P}_{\ell M}$, proportional to the off-diagonal correlators ${}^{XY}{\cal A}^{M}_{\ell\ell\pm1}$, are shown on Fig.~\ref{Fig3}. However, the shear induces a $B$-mode spectrum whose amplitude is about $(\sigma/H)^2$ lower than for the one for the $E$-mode in the most optimistic model ($B$). We can compare our results to the bounds set by the CFHTLS survey~\cite{kitching}. Unfortunately, CFHTLS covers 4 fields of typical size 50~sq. deg. so that the largest scale with a sufficiently good signal-to-noise ratio is of the order of $\ell\sim2000$, far beyond the linear regime.  $B$-modes will be generated from the non-linear dynamics and it is safer to rely on the $EB$ cross-correlation. To get a rough idea, though, we use the values at $\ell=2000$, for which $\ell^2C_\ell^{EE}\sim 10^{-6} $ and $\ell^2C_\ell^{EB}<4\times10^{-7}$. Indeed this estimation has to be taken with a grain of salt given the fact that i) there is a large scatter in Fig. 6 and 7 of Kitching et al. (2014) and ii) these observations are not in the linear regime and there is no unambiguous way to scale these observations to lower $\ell$. 
We can thus set the bound $(\sigma/H)_0 \lesssim 0.4$ from CFHTLS. However, Euclid shall probe scales up to 100 deg., deep in the linear regime, with a typical improvement of a factor 50~\cite{privatecommun}. This would  translate to a sensitivity of order $(\sigma/H)_0 \lesssim 0.4/50\sim1\%$ for the shear. This estimate indicates that weak-lensing could be a powerful tool to constrain a late time anisotropy. In the meantime, experiments such as DES will allow us to forecast more precisely the power of Euclid.  It demonstrates that in principle one can reconstruct the principal axis of expansion from observations. We thus want to draw the attention on the importance of these estimators and their measurements.

{\it Conclusion}: this letter emphasises the specific signatures of an anisotropic expansion on weak-lensing, as first pointed out in Ref.~\cite{pup}. Following our formalism detailed in Ref.~\cite{newppu} (where all the technical details can be found), we have focused on two phenomenological anisotropic models and computed the angular power spectra of the $E$- and $B$-modes, as well as the five non-vanishing off-diagonal correlators. These are {\em new} observables that we think must be measured in future surveys. These measurements can be combined easily with the Hubble diagram since the Jacobi matrix can be determined analytically at background level~\cite{Fleury2014}. Let us emphasize that the off-diagonal correlations with the polarisation can also be applied to the analysis of the CMB, hence generalising easily those built from the temperature alone~\cite{nondiag-coor}.

Our analysis demonstrates that future surveys, and in particular Euclid, can set strong bounds on the anisotropy of the Hubble flow, typically at the level of $(\sigma/H)_0 \leq 1\%$, but one needs a detailed analysis of the signal-to-noise ratio to confirm this number, which for now has to be taken
as an indication. This is a new and efficient method and the estimators built from the off-diagonal correlators can be used to reconstruct the proper axis of the expansion.

\end{document}